\documentstyle[fullpage,fleqn]{article}
\font\fonta=cmr12 scaled\magstep1
\font\fontb=cmr12 scaled\magstep2
\font\fontc=cmr12 scaled\magstep1

\begin{document}
{
\begin{flushright}{TMU-NT931001\\October 1993}
\end{flushright}
}
\vglue 0.5cm
\large
\baselineskip=1cm
\begin{center}{\fontb Diquark Correlations\\in the Nucleon Structure 
Function 
               } \\
\vglue 2cm
{\fonta Katsuhiko Suzuki, Takayuki Shigetani and  Hiroshi Toki\footnote
{{\fontc also RIKEN, Wako, Saitama, 351, Japan}}\\
{\fontc \em Department of Physics, Tokyo Metropolitan University}\\
{\fontc \em Hachiohji,Tokyo 192, Japan}}
\end{center}
\vglue 2.7cm
\baselineskip=0.8cm
\noindent
Abstract : The nucleon structure functions are studied with a special 
emphasis on diquark correlations inside the nucleon.  Assuming a 
phenomenological diquark-quark model of the nucleon at the low energy 
hadronic scale, we calculate the leading twist contributions to the 
nucleon structure functions.  We take the Nambu and Jona-Lasinio model 
for the calculation of the quark distributions in the diquarks, and use 
the convolution method for the diquark scattering process.  The 
resulting quark distributions are evolved to the experimental momentum 
scale using the perturbative QCD.  We discuss the flavor dependence of 
the nucleon structure functions and the importance of the quark-quark 
correlations for the understanding of the nucleon structure.
\newpage
\noindent
{\fonta {\bf 1 Introduction}}
\baselineskip=0.8cm

Recent measurements of deep inelastic scattering show clear flavor 
dependences of the nucleon structure functions[1-3].  One example is 
the ratio of the neutron to proton structure functions 
\(F_{2}^{n}(x)/F_{2}^{p}(x)\)  which shows a large deviation from the 
naive quark-parton model prediction, which is 2/3.  In addition, the 
recent NMC data shows that the difference between the \(Q^2\) 
dependence of \(F_{2}^{p}(x)\) and \(F_{2}^{n}(x)\) is substantially 
larger than the perturbative QCD prediction\cite{NMC}.

On the other hand, the diquark-quark model of the nucleon has attracted 
considerable interests.  The idea of diquarks, i.e. correlated states 
of two quarks, is introduced phenomenologically to explain the scaling 
violation of the nucleon structure functions, which may be caused by 
the non-perturbative diquark correlations\cite{Land}.  The diquark 
model also reproduces the proton-neutron differences of the structure 
functions\cite{Stock}.  Such a diquark structure is due to the 
quark-quark correlations which depend on its flavor structure, and may 
produce an asymmetry of the momentum distribution in the nucleon.  In 
fact, such asymmetric momentum distributions are found in the 
calculations of the QCD sum rule\cite{QSR1,QSR2}, where the most part 
of the proton momentum is carried by the u-quark with its spin 
directed along the proton spin and the remaining small part is carried 
by the u-d quarks with combined spin-0.  Additionally, the elastic 
form factor of the nucleon requires an asymmetric momentum 
distributions, which indicates the diquark clustering inside the 
nucleon\cite{Stoll}.  This model is also supported by the study of 
several high energy scattering processes\cite{Hirose,Diquarks}.

In the low energy side of the hadron physics, the diquark-quark 
structure of baryons also plays various 
roles\cite{Diquarks,Ida,Lichtenberg}.  This picture was first 
introduced to explain the universality of the Regge 
trajectories\cite{Ida}.  In particular, a recent work of Stech and his 
collaborators showed that the diquark correlations, which are the 
strongest in the scalar diquark channel, can account for the 
\({\Delta}I=1/2\) problem of weak non-leptonic transitions, and 
reproduce the experimental data very well\cite{Stech,SuzukiW}.  Hence, 
it is of our great interest to study the flavor dependence of the 
nucleon structure in terms of the diquark-quark model as a successful 
low energy model of baryons.  We note that such an approach naturally 
reproduces the asymmetry of momentum distributions in the nucleon 
obtained by the QCD sum rule.

Several theoretical calculations of the nucleon structure function were 
made in terms of the low energy quark models[15-20].  The structure 
function at the low energy scale \(Q^{2} = Q_{o}^{2}\), where the 
phenomenological quark model is supposed to work, is obtained by 
calculating the twist-2 matrix elements, since these matrix elements 
are dominant over higher twist terms in the Bjorken limit.  The 
resulting structure functions are evolved to the experimental scale 
with the help of the perturbative QCD.  Thus, one can compare them 
with experimental data.  Here, we use the same method to evaluate quark 
distributions within the diquark model.

Recently, the importance of the quark-diquark intermediate state in the 
calculation of the nucleon structure function was discussed within the 
MIT bag model\cite{Close} and the quark-diquark model\cite{MM}.  It was 
pointed out that the mass difference between the scalar($0^{+}$) and 
the axial-vector($1^{+}$) diquarks is crucial to account for the flavor 
dependence of quark distributions, since this mass difference produces 
the dominance of u-quark distribution in the proton at large Bjorken 
\(x\).  In such works, the diquarks are treated only as a spectator.  
However, if the two quark correlation is so strong as obtained in the 
QCD sum rule, we should take into account the diquark scattering 
process explicitly with the other quark being a spectator.

In this paper, we shall calculate the nucleon structure function in the 
framework of the $SU(6)$ diquark-quark model.  Thus, both quark and 
diquark scattering processes contribute to the structure function, 
which are illustrated in Fig.1(a) and (b).  The process of Fig.1(a), 
in which a quark is struck out by the virtual photon with the residual 
diquark being a spectator, is evaluated by using the standard method in 
the impulse approximation\cite{MM}.  For the diquark part, we use the 
convolution technique\cite{Sullivan,Convolution} to calculate 
contributions of the diquark process shown in Fig.1(b).  We obtain the 
diquark terms as the convolution of quark distributions in diquarks 
with the diquark distributions in the nucleon.  This diagram represents 
the leading twist contributions of diquarks, in which diquarks break up 
completely after absorbing the virtual photon\cite{Land}.  Hence, the 
resulting quark distributions scale; i.e. there is no \(Q^2\) 
dependence without the QCD radiative correction.  Here, we shall not 
calculate the twist-4 non-scaling contributions of diquarks, which are 
under consideration\cite{SuzukiPre}.

\vglue 1.0cm

\noindent
{\fonta {\bf 2. Nucleon Structure Function in the Diquark model}}
\baselineskip=0.8cm

The $SU(6)$ diquark-quark wave function of proton is written 
as\cite{Lichtenberg},
%
%
\vglue 0.1cm
\baselineskip=0.8cm
\begin{eqnarray}
|p\uparrow >={1 \over {\sqrt {18}}}[ 3S(ud)u\uparrow + 
\, 2A(uu)^{+}d\downarrow 
&-&\sqrt 2A(uu)^{o}d\uparrow \nonumber\\
&-&\sqrt 2A(ud)^{+}u\downarrow + \, A(ud)^{o}u\uparrow ]
\label{WF}
\end{eqnarray}
\vglue 0.3cm
\noindent
where \(S(ij)\) denotes the scalar (spin singlet; $0^{+}$) diquark with 
the flavor content \(i\) and \(j\), and \(A(ij)^\kappa\) the 
axial-vector (spin triplet; $1^{+}$) diquark with the helicity 
\(\kappa\).  In this model, both the quark and the diquark scattering 
processes contribute to the hadronic tensor.  Using (\ref{WF}), the 
proton and the neutron structure functions are given by,
%
%
\vglue 0.1cm
\baselineskip=0.8cm
{
\setcounter{enumi}{\value{equation}}
\addtocounter{enumi}{1}
\setcounter{equation}{0}
\renewcommand{\theequation}{\theenumi-\alph{equation}} 
\begin{eqnarray}
F_{2}^{p}(x)={2 \over 9}q^{S}(x)+{1 \over 9}q^{V}(x)+
{5 \over {18}}Q_{D}^{S}(x)+{7 \over {18}} Q_{D}^{V}(x)
\label{f2p}
\end{eqnarray}
\begin{eqnarray}
F_{2}^{n}(x)={1 \over {18}}q^{S}(x)+{1 \over 6}q^{V}(x)+
{5 \over {18}}Q_{D}^{S}(x) +{1 \over 6}Q_{D}^{V}(x)
\label{f2n}
\end{eqnarray}
\setcounter{equation}{\value{enumi}}
}
\vglue 0.3cm
\noindent
where \(q^S\) and \(q^V\) are the quark distributions with the residual 
diquarks being the scalar and the axial-vector diquarks, respectively.  
\(Q_D^S\) and \(Q_D^V\) are the quark distributions obtained by the 
scalar and the axial-vector diquark scattering processes, 
respectively.  We can also express the valence distributions of the u 
and d quarks in the proton as,
%
%
\vglue 0.1cm
{
\setcounter{enumi}{\value{equation}}
\addtocounter{enumi}{1}
\setcounter{equation}{0}
\renewcommand{\theequation}{\theenumi-\alph{equation}} 
\begin{eqnarray}
u(x)={1 \over 2}q^{S}(x)+{1 \over 6}q^{V}(x)+{1 \over 2}Q_{D}^{S}(x)+
{5 \over 6}Q_{D}^{V}(x)
\label{uval}
\end{eqnarray}
\begin{eqnarray}
d(x)={1 \over 3}q^{V}(x)+{1 \over 2}Q_{D}^{S}(x)+{1 \over 6}Q_{D}^{V}(x)
\label{dval}
\end{eqnarray}
\setcounter{equation}{\value{enumi}}
}
\vglue 0.3cm
\noindent
Note that \(q^{S}(x)\) term contributes only to the u-quark 
distribution.  The flavor dependence of the structure functions arises 
from the difference of the distribution contents in this model.  
We also note that the definition (\ref{f2p},\ref{f2n}), 
(\ref{uval},\ref{dval}) is different from that of the previous 
work\cite{Stock,Diquark}, in which the diquarks are assumed to be 
point-like.  We consider the scaling contributions of diquark 
processes, where the diquarks are broken up completely by the virtual 
photon. 
\vglue 1.0cm

\noindent
{\fonta {\bf 3 Nambu and Jona-Lasinio model}}
\baselineskip=0.8cm

We use the Nambu and Jona-Lasinio (NJL) model to take into account the 
quark correlations, where the chiral invariance is the main 
ingredient\cite{NJL}.  Recently, the NJL model is studied in various 
subjects of hadron physics as a low energy effective theory of 
QCD\cite{Vogl}.  This model describes the \(SU(3)_f\)  meson properties 
very well with the parameters fixed by the pion and the kaon 
properties, in spite of the lack of confinement.  The \(SU(3)\)  NJL 
lagrangian for mesons is given by\cite{Vogl},
%
%
\vglue 0.1cm
\begin{eqnarray}
{\cal L}_{NJL}&=&{\cal L}_{o} + {\cal L}_{M}\nonumber\\
{\cal L}_{o}  &=& \bar{\psi}(i\gamma^{\mu}\partial_{\mu} - m_{o})
\psi\nonumber\\
{\cal L}_{M}  &=& G_{S}^{M}[(\bar{\psi}t_{a}\psi)^2
                       +(\bar{\psi}t_{a}i\gamma^{5}\psi)^2]\nonumber\\
              & &-G_{V}^{M}[(\bar{\psi}t_{a}\gamma_{\mu}\psi)^2
                       +(\bar{\psi}t_{a}\gamma_{\mu}\gamma^{5}\psi)^2]
\end{eqnarray}
\vglue 0.3cm
\noindent
Here \(\psi\) denotes quark fields and \(m_0\) their bare masses.  
\(t_a\) are the flavor \(SU(3)_f\) operators with the normalization 
\(tr(t_{a}t_{b})=\delta_{ab}/2\), and \(G_{S}^{M}\), \(G_{V}^{M}\) are 
the coupling constants.  In this model, the quarks acquire the 
constituent masses dynamically due to the spontaneous breakdown of the 
chiral symmetry.  The meson masses are obtained by solving the 
Bethe-Salpeter (BS) equation.

On the other hand, the NJL lagrangian for the quark-quark interaction 
at the Fierz transformed level is given by[26,27],
%
%
\begin{eqnarray}
 {\cal L}_{D}  &=& G_{S}^{D}[(\bar{\psi}t_{a}c_{A}\psi_{c})
 ({\bar{\psi}}_{c}t_{a}c_{A}\psi)
+(\bar{\psi}t_{a}c_{A}i\gamma^{5}\psi_{c})
({\bar{\psi}}_{c}t_{a}c_{A}i\gamma^{5}\psi)]\nonumber\\
&-&G_{V}^{D}[(\bar{\psi}t_{s}c_{A}\gamma^{\mu}\psi_{c})
({\bar{\psi}}_{c}t_{s}c_{A}\gamma_{\mu}\psi)\nonumber\\
&+&(\bar{\psi}t_{a}c_{A}\gamma^{\mu}\gamma^{5}\psi_{c})
({\bar{\psi}}_{c}t_{a}c_{A}\gamma_{\mu}\gamma^{5}\psi)]
\end{eqnarray}
\vglue 0.3cm
\noindent
Here, \(\psi_{c}=C\bar{\psi}^{T}\) are the charge conjugate quark 
fields and \(c_A\) the antitriplet color \(SU(3)\) operators with the 
normalization \(tr(c_{i}c_{j})=\delta_{ij}/2\), where \(A\) runs over 
only \(A=\) 2, 5 and 7.  The flavor suffix \(a\) runs over \(a\) = 2, 5 
and 7 for pseudoscalar, scalar and vector diquark terms, while it runs 
over \(s\) = 0, 1, 3, 4, 6 and 8 for the axial-vector diquark term due 
to the antisymmetrization of diquarks.  Note that the presence of the 
charge conjugation operator produce an additional minus sign to the 
parity of diquarks.  We get the masses and the wave functions of 
diquarks by solving the quark-quark BS equations as the case for 
mesons.  Recently, the diquark structure is extensively studied in the 
framework of the NJL model\cite{Vogl}.  We find that the structure of 
diquarks is quite similar with that of mesons with the corresponding 
quantum numbers.  In particular, the scalar diquark 
(\(\bar{\psi}_{c}i\gamma_{5}\psi\)), which corresponds to the pion 
(\(\bar{\psi}i\gamma_{5}\psi\)), shows a strong quark-quark 
correlation, whereas the axial-vector diquark 
(\(\bar{\psi}_{c}\gamma_{\mu}\psi\)), which corresponds to the rho 
meson (\(\bar{\psi}\gamma_{\mu}\psi\)), is a weakly bound quark-quark 
state.  If the coupling constants satisfy the condition 
\(G_{S}^{M}=G_{S}^{D}\) and \(G_{V}^{M}=G_{V}^{D}\), the diquark mass 
is the same as the corresponding meson mass, which is so-called the 
Pauli-G\"{u}rsey symmetry between mesons and 
diquarks\cite{SuzukiW,Vogl,SuzukiPG}.  Note that the strong diquark 
correlation in the scalar channel causes the large enhancement of the 
\({\Delta}I=1/2\) transition matrix elements of the hyperon non-
leptonic weak decays\cite{Stech,SuzukiW}.

The nucleon properties such as the mass and the wave function within 
the NJL model should be obtained by solving the relativistic three body 
problems, and such an effort is being made with the relativistic 
Faddeev method\cite{Faddev}.  It seems, however, necessary to 
incorporate confinement for a successful description of baryons, which 
is absent in the NJL model.  Therefore, we simply assume the 
diquark-quark model for the nucleon, and consider the case of 
diquark-quark vertex being scalar\cite{MM};
%
%
\begin{eqnarray}
\Gamma _{Dq}(p^2)\propto {\bf1}\cdot \phi (p^2)
\end{eqnarray}
\vglue 0.3cm
\noindent
where \(\bf{1}\) is the unit matrix in the Dirac space and 
\(\phi(p^2)\) is a regularization dependent function to be specified 
later.  The diquark properties are obtained by solving the NJL model, 
which tells us the information of the quark-quark correlations.

\vglue 1.0cm
\noindent
{\fonta {\bf 4. Calculations of Hadronic Tensor}}
\baselineskip=0.8cm

First, we evaluate contributions shown in Fig.1(a), where a quark is 
struck out by the virtual photon with the residual diquark being a 
spectator.  This part can be calculated following the work of Meyer 
and Mulders\cite{MM}.   We use the impulse approximation, and thus the 
hadronic tensor is represented by an incoherent sum of various 
processes.  We define the constituent quark mass \(m\) and the diquark 
mass \(m_D\) inside the nucleon, though the diquark and the quark are 
not the eigenstates of QCD.  Their values are obtained within the NJL 
model.  The calculation of the hadronic tensor in the Bjorken limit 
yields\cite{MM},
\vglue 0.15cm
\begin{eqnarray}
q^D(x)&=&\int {{{d^4p} \over {(2\pi )^4}}}{{\phi ^2} 
\over {(p^2-m^2)^2}}2\pi \delta (p_{2}^2-m_D^2)
\theta (p_2^0)\nonumber\\
& &                \hspace{2cm} \times \, {1 \over {2M\nu }}
Tr[(p\kern -1.7mm /+m)\gamma ^+(p\kern -1.7mm /+m)(P\kern -2.5mm /+m)
]\nonumber\\
&=&\int_{p_{E\min }^2}^\infty  {{{dp_E^2} \over {8{\pi}^2}}}
{{\phi ^2(p_E^2)} \over {(p_E^2+m^2)^2}}
[x(M^2+2mM-m_D^2)+m^2+(1-x)p_E^2]
\label{qx}
\end{eqnarray}
\vglue 0.3cm
\noindent
where 
\begin{eqnarray*}
p_{Emin}^{2} = {x \over 1-x}m_D^2 - xM^2 ,
\end{eqnarray*}
\noindent
and $M$ is the nucleon mass.  Here, $P$ is the proton momentum, and 
$p$ and $p_2$ are momenta of the struck quark and the spectator 
diquark, respectively.

As for the diquark process shown in Fig.1(b), we fold the quark 
distributions in diquarks with the diquark distributions in the 
nucleon.  We use the NJL model for the calculation of the quark 
distributions in the diquarks.  Note that these contributions to the 
structure functions scale.  The approximation of the convolution method 
may be good for the scalar diquark process, since the scalar diquark is 
a small object due to its strong correlation.  In fact, the NJL model 
calculation gives a small radius $\leq$ 0.4fm.  Phenomenological 
estimates of the scalar diquark radius also provides small 
value\cite{Stock,Diquarks}.  But the size of the axial-vector diquark 
is comparable with that of the nucleon, thus the convolution is not 
suitable for the calculation of the axial-vector diquark process.   
Assuming the quark-diquark vertex to be scalar as mentioned above, we 
obtain the diquark term as the convolution integral\cite{Convolution};
\vglue 0.05cm
\begin{eqnarray}
Q_D^i(x)=\int_x^1 {dx/y}\, F_{D/N}(y)\, D_{Di}(x/y)
\label{eqq}
\end{eqnarray}
\vglue 0.3cm
\noindent
$F_{D/N}(y)$ is a probability to find a diquark in the nucleon with the 
light-cone momentum fraction $y$.  This part is evaluated in terms of 
the scalar vertex (6);
%
%
\vglue 0.15cm
\begin{eqnarray}
F_{D/N}^{}(y)&=&y\int_{}^{} {{{d^4k} \over {(2\pi )^4}}}\phi^2 
(k^2){{2[M^2+mM-P\cdot k]} \over {(k^2-m_D^2)^2}}\nonumber\\
& &         \hspace{2cm}    \times \,  2\pi \delta (k_2^2-m^2)
\delta (y-{{k^+} \over M})\theta (k^+)\theta (M-k^+)\nonumber\\
&=&y\int_{t\min }^{} {{{dt} \over {16\pi ^2}}}\phi (t)^2{{[(m_D+M)^2+t]}
\over {(t+m_D^2)^2}} \, ,
\label{qdx}
\end{eqnarray}
\vglue 0.3cm
\noindent
where
\begin{eqnarray*}
t_{min}={y \over {1-y}}m^{2} - yM^{2} .
\end{eqnarray*}
\noindent
Here, the diquark carries the momentum $q$ and the residual quark 
$q_2$.   $D_{Di}(x)$ is the quark distribution of the leading twist 
diquark structure function.  For the calculation of $D_{Di}(x)$, we 
use the same procedure as done in the meson case within the NJL 
model[29].  For example, the quark distributions in the scalar diquark 
is given by,
%
%
\vglue 0.15cm
\begin{eqnarray}
D_{S}(x)  &{\propto}&   g_{Sqq}^{2}\int\! d\mu^{2}_{E}
[\frac{1}{\mu^{2}_{E}+m^{2}}
+x\frac{m_{D}^{2}}{(\mu^{2}_{E}+m^{2})^{2}}] f(\mu_E^2) \nonumber\\
& &\hspace{2cm} \times \, 
\theta (m_{D}^{2}x(1-x)-xm^{2}+(1-x)\mu^{2}_{E}) \, .
\label{dx}
\end{eqnarray}
\vglue 0.3cm
\noindent
$g_{Sqq}$ is the diquark-quark-quark coupling constant obtained by the 
NJL model calculation.  $f({\mu}_{E}^{2})$ is the regularization 
function.  Note that the form of (\ref{dx}) is the same as the quark 
distribution of the pion obtained in ref. \cite{Shige1}, since the 
diquark structure is the same as that of the corresponding meson in 
the NJL model due to the relation\cite{Vogl,SuzukiPG};

\hspace{0.5cm}	\(C S_{F}^{T}(q) C^{-1} = S_{F}(-q)\)

\noindent
where $C$ is the charge conjugation operator and $S_F$ the quark 
propagator.  The structure function of the axial-vector diquark is also 
calculated in the similar manner\cite{Shige2}.  Thus, we get the 
convolution of the axial-vector diquark term.

The integrals (\ref{qx}), (\ref{qdx}), and (\ref{dx}) go to infinity 
due to the non-renormalizability of the NJL model.  To avoid the 
divergence, we introduce the regularization function $f(p^{2})$ in the 
Euclidean space.  There exists an difficulty in the introduction of the 
standard NJL cutoff for DIS process\cite{Sea1,Sea2}.  Here, we use the 
Fermi distribution type cut-off function, which is consistent to the 
usual sharp cut-off method except for large momentum. 
\begin{eqnarray}
f(t)={1 \over {1+\exp [(t-\Lambda ^2)/a]}}=\phi^2 (t)
\end{eqnarray}
\vglue 0.3cm
\noindent
Here, we use ${\Lambda}{\sim}1GeV$ and $a\sim0.1GeV^2$ to reproduce the 
meson properties. We examine also other plausible form like the 
exponential cut-off for completeness.  We obtain similar quark 
distributions except for the shape around $x\sim1$, which depends on 
the high momentum behavior of the cut-off scheme.

\vglue 1.0cm
\noindent
{\fonta {\bf 5. Results and Discussions}}
\baselineskip=0.8cm

We present here the calculated result on the structure function.  
Concerning the NJL model parameters, we take the current quark mass 
\(m_o=5.5MeV\) as a semi-empirical value, and \({G_S^M}{\Lambda}^2=
2{G_V^M}{\Lambda}^2=22.2,\, \Lambda=860MeV,\, a=0.15GeV^{2}\) which are 
fixed by the pion properties. For the diquark part, we treat the 
diquark masses as free parameters, and study their dependence of the 
structure function.  By the analysis of the N-${\Delta}$ mass splitting 
in the one-gluon exchange picture, the scalar diquark mass \(m_S\) was 
assumed to be \(575MeV\)\cite{OGE}, and the axial-vector diquark mass 
\(m_A\) to be \(200MeV\) higher than the scalar diquark 
mass\cite{Close,MM}.  Much larger scalar-vector diquark mass difference 
was obtained in the instanton liquid model\cite{Shuryak} or other 
models\cite{Diquarks,Vogl}.  Here, we assume \(m_A=775MeV\), which is 
consistent to the above arguments and comparable with the vector meson 
mass like ${\rho}$.  Then, we investigate the scalar diquark mass 
dependence of the results.  The quark correlation affects the quark 
distributions through the diquark masses.

We first show the quark distributions of various processes at the low 
energy scale in Fig.2.  These distributions represent their strength 
of the correlation.  The quark distribution \(q^{S}(x)\) (solid curve), 
where the scalar diquark is a spectator, has a sharp distribution 
peaked at \(x\sim0.5\), because of the weakness of the interaction 
between the scalar diquark and the third quark due to the strong 
binding of the diquark in the scalar channel.  \(q^{V}(x)\) 
(dashed curve) shows slightly a broader distribution.  
\(q^{S}(x)\) and \(q^{V}(x)\) are almost unchanged from the results of 
ref. \cite{MM}.  In addition, we have calculated newly the 
contributions from diquark process (Fig.1(b)).  The momentum of two 
quarks in the scalar diquark is, compared to \(q^{S}(x)\), peaked at 
small \(x; x\sim0.25\), as shown by dotted line.  Using these 
distributions, we can obtain the momentum ratio carried by each quark.  
For the scalar channel, we find at \(m_{S}=575MeV\)

	\(<xq^{S}> : <xQ_D^{S}>\sim  2.2 : 1\) ,

\noindent
where \(<xq>= \int_0^1 {dx} xq(x)\).  This means that the u-quark with 
the scalar diquark being a spectator carries the largest part of the 
proton momentum, and the remaining small part is carried by the u-d 
scalar diquark due to the strong correlation in the scalar channel.  
This result is consistent with the QCD sum rule\cite{QSR1,QSR2}.  If 
the diquark correlation becomes weak, we find that this ratio is close 
to 1.  For example, $<xq^{S}> : <xQ_D^{S}>\sim 1.3 : 1$ at 
$m_{S}=775MeV$.  In the case of axial-vector diquark process, we obtain

	 \(<xq^{V}> : <xQ_D^V>\sim 1.3 : 1\).

\noindent
Thus, both two quarks in the axial-vector diquark ($Q_{D}^{V}(x)$: 
dash-dotted curve) and the third quark ($q_{V}(x)$) have almost the 
same amount of the proton momentum.  This fact reflects the weak 
correlation of quarks in the axial-vector diquark channel.  Comparing 
$Q_{D}^{S}(x)$ with $Q_{D}^{V}(x)$, the difference of both 
distributions is not so evident.  Originally, the shapes of 
distributions in the scalar and axial-vector diquarks are quite 
different, as $\pi$ and $\rho$ in the meson case\cite{Shige2}.  
However, such a difference is washed out after the convolution with the 
entire diquark motion in the proton.

We take the low energy model scale \(Q_{o}^{2}=0.2GeV^2\), which is 
used in ref. \cite{Gluck}.  At this scale, the running coupling 
constant is still small; \(\alpha_{s}(Q^{2}_{o})/\pi\sim0.4\).  
Thus, the inclusion of the second order QCD corrections gives a small 
change for the \(Q^{2}\) evolution from our result within 
10\%\cite{Gluck}.  We use the first order Altarelli-Parisi 
equation\cite{AP} with the \({\Lambda}_{QCD}=250MeV\), in order to 
compare our result with experiment.  For the sea quark and the gluon 
parts, we simply assume that these distributions vanish at 
\(Q^{2}=Q_{o}^{2}\), and are generated by the QCD evolution process.

The proton and the neutron structure functions \(F_{2}^{p}(x)\) and 
\(F_{2}^{n}(x)\) thus obtained at \(Q^{2}=15GeV^2\) are shown in Fig.3 
with experimental data\cite{EMC}, where we use 
\(m_{A}-m_{S}= 200MeV\).  Our result shows a reasonable agreement with 
data for \(x>0.4\), and is somewhat different for small \(x\) value.  
One of the possible solutions for this disagreement is to take into 
account the sea quark degrees of freedom explicitly at the low energy 
scale.  The inclusion of the \(\bar{q}-4q\) or \(q-3q\bar{q}\)  
intermediate state at the model scale may change the shape of 
\(F_{2}(x)\) at small \(x\) as introduced in ref. \cite{Thomas}.  
Comparing our calculated results with the ones of the MIT bag 
model\cite{Miller,Thomas} or the diquark spectator model\cite{MM}, our 
structure functions are distributed over much wider range of the 
Bjorken \(x\).   This fact is the consequence of the strong diquark 
correlation.   We also show in Fig.3 the proton structure function 
without the diquark scattering part (Fig.1(a) only), in which we use 
\(Q_{o}^{2}=0.2GeV^2\).  This result is essentially the same as that of 
the previous model calculation without the diquark process[16-20].  
This structure function has a sharp peak around \(x\sim0.3\), and 
disagrees with experiment.  In order to reproduce the absolute values 
and the peak position of the experimental data, one must take a smaller 
value for the model scale \(Q_{o}^{2}\).  In fact, a much smaller value 
for the low energy scale is used in the previous 
calculations\cite{Thomas,Models,MM} so as to obtain a better agreement 
with experiment.

We would like to discuss the dependence of the structure function on 
the low energy scale \(Q_o^2\).  The nucleon structure functions are 
shown in Fig.4, with the choice of the low energy scale 
\(Q_{o}^{2}=0.125GeV^2\).  The running coupling constant \({\alpha}_s\) 
at this scale is about two times larger than that for 
\(Q_{o}^{2}=0.2GeV^2\).  As compared with the result in Fig.3, the 
calculated structure functions are in a good agreement with experiment, 
especially at low \(x\).  However, the use of the perturbative QCD 
becomes less reliable at such a low \(Q_{o}^{2}\).

We discuss how the flavor structure of the structure function depends 
on the diquark correlation.  The ratio \(F_{2}^{n}(x)/F_{2}^{p}(x)\) is 
shown in Fig.5.  Our result is in good agreement with 
experiment\cite{EMC,NMC,BCDMS}, in the case \(m_{A}-m_{S}= 200MeV\).  
This is due to the dominance of u-quark distributions, namely 
\(q^{S}(x)\), at large \(x\), which is caused by the asymmetric 
momentum distributions of quarks and diquarks.  In the middle \(x\) 
range, \(x\sim0.4\), the resulting ratio is close to 2/3, since this 
part of distributions is dominated by the diquark parts, i.e. 
\(Q_{D}^{S}(x)\) and \(Q_{D}^{V}(x)\).  We find that our calculated 
result is somewhat smaller than experiment for small \(x\).  The 
inclusion of sea quark at the low energy scale enhances this ratio in 
the small \(x\) region, and may resolve this discrepancy.  We also show 
in Fig.5 the case \(m_{A}-m_{S}=0, \, 100, \, 275MeV\).  If we take 
the same values for the scalar and the axial-vector diquark masses, the 
ratio is close to 2/3.  We also show the difference 
\(F_{2}^{p}(x)-F_{2}^{n}(x)\) in Fig.6.  The peak position of our 
result is consistent with experiment, but the absolute value is rather 
large.  This behavior was found in the previous calculations in terms 
of other models.

\vglue 1cm

\noindent
{\fonta {\bf 6. Conclusion}}
\baselineskip=0.8cm

In conclusion, we have studied the flavor structure of quark 
distributions based on the phenomenological diquark model.  This model 
reproduces the asymmetric momentum distributions in the nucleon 
obtained by the QCD sum rule.  The neutron to proton structure function 
ratio \(F_{2}^{n}(x)/F_{2}^{p}(x)\) shows a good agreement with the 
available DIS data.  The agreement is lost as the diquark correlation 
becomes weak.  Our result indicates that the quark correlation, 
especially in the scalar channel, is important to understand the DIS 
data, which is also crucial for the low energy hadron properties such 
as the \({\Delta}I=1/2\) non-leptonic weak transition.  Although we 
consider the twist-2 contributions of the structure function in this 
paper, twist-4 pieces of them give us much information about the 
quark-quark correlation\cite{Hatsuda}.  The diquark-like correlation 
may be important to study such higher twist effects.  Work along this 
line is in progress.

\newpage
\noindent
{\bf Figure Caption\/}
\vglue 0.5cm

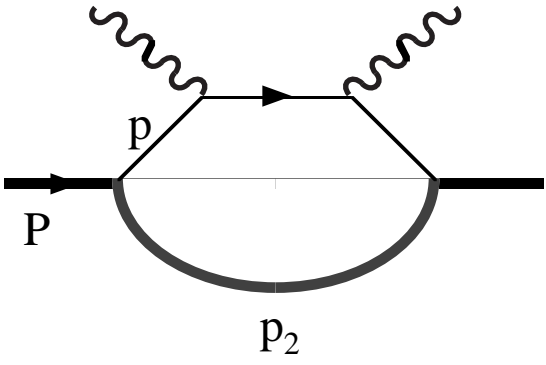
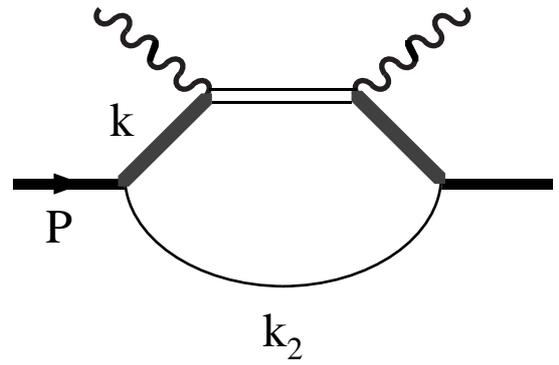
Fig. 1	: The forward scattering amplitude ("handbag diagram") of the 
nucleon.  The thin solid line represent the quark, and the shaded line 
the diquark.  The nucleon and the virtual photon are depicted by the 
thick solid and wavy lines.  For details and notation, see text.
\vglue 0.5cm

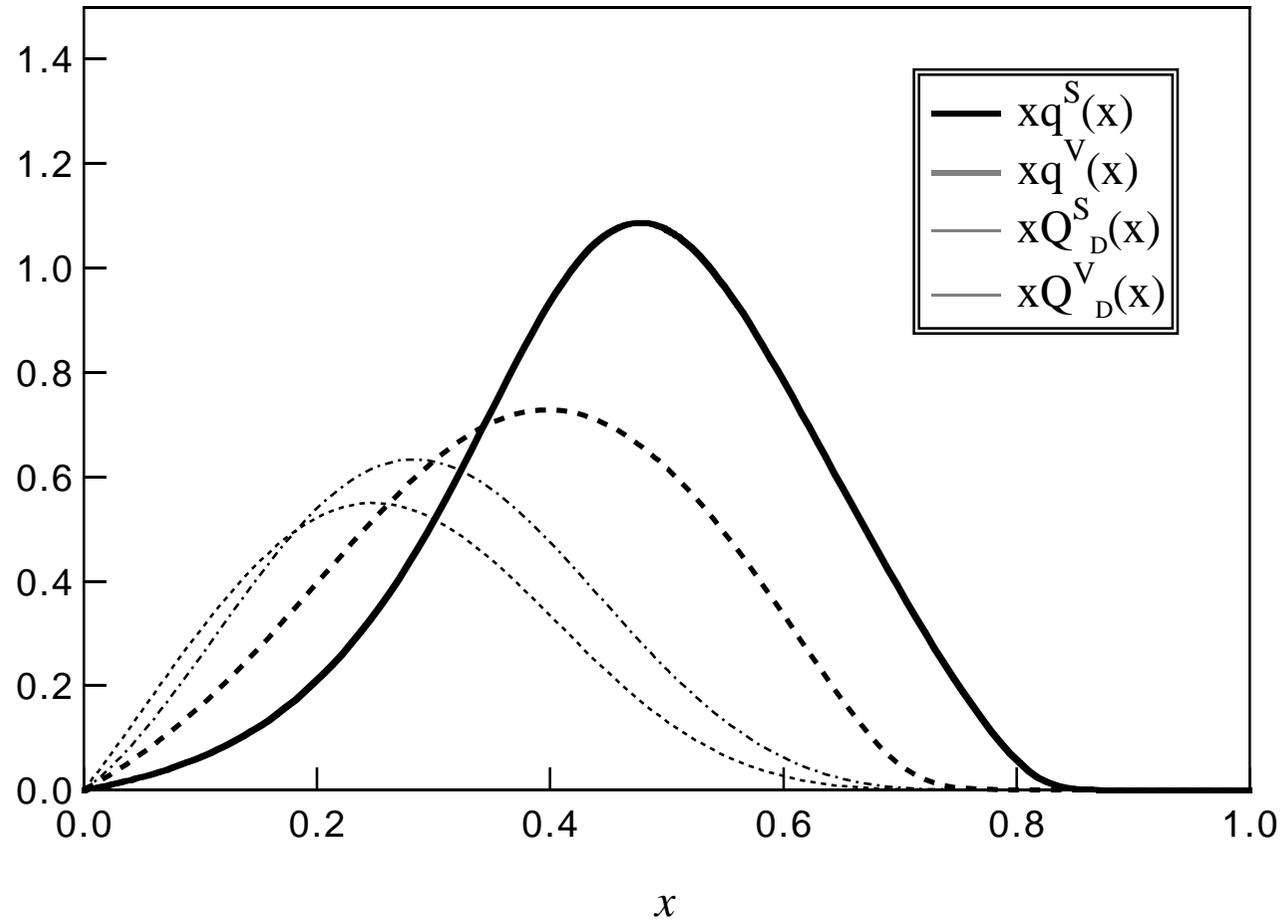
Fig. 2	: The quark distributions of the nucleon at the low energy 
scale \(Q^{2}=Q_{o}^{2}\) as a function of the Bjorken \(x\). The solid 
and dashed curves denote the quark distributions obtained by the 
calculation of Fig.1(a) with the residual diquarks being scalar or 
axial-vector, respectively.  The dotted and dash-dotted curves 
represent the distributions of the scalar and axial-vector diquark 
scattering processes (Fig.1(b)).
\vglue 0.5cm

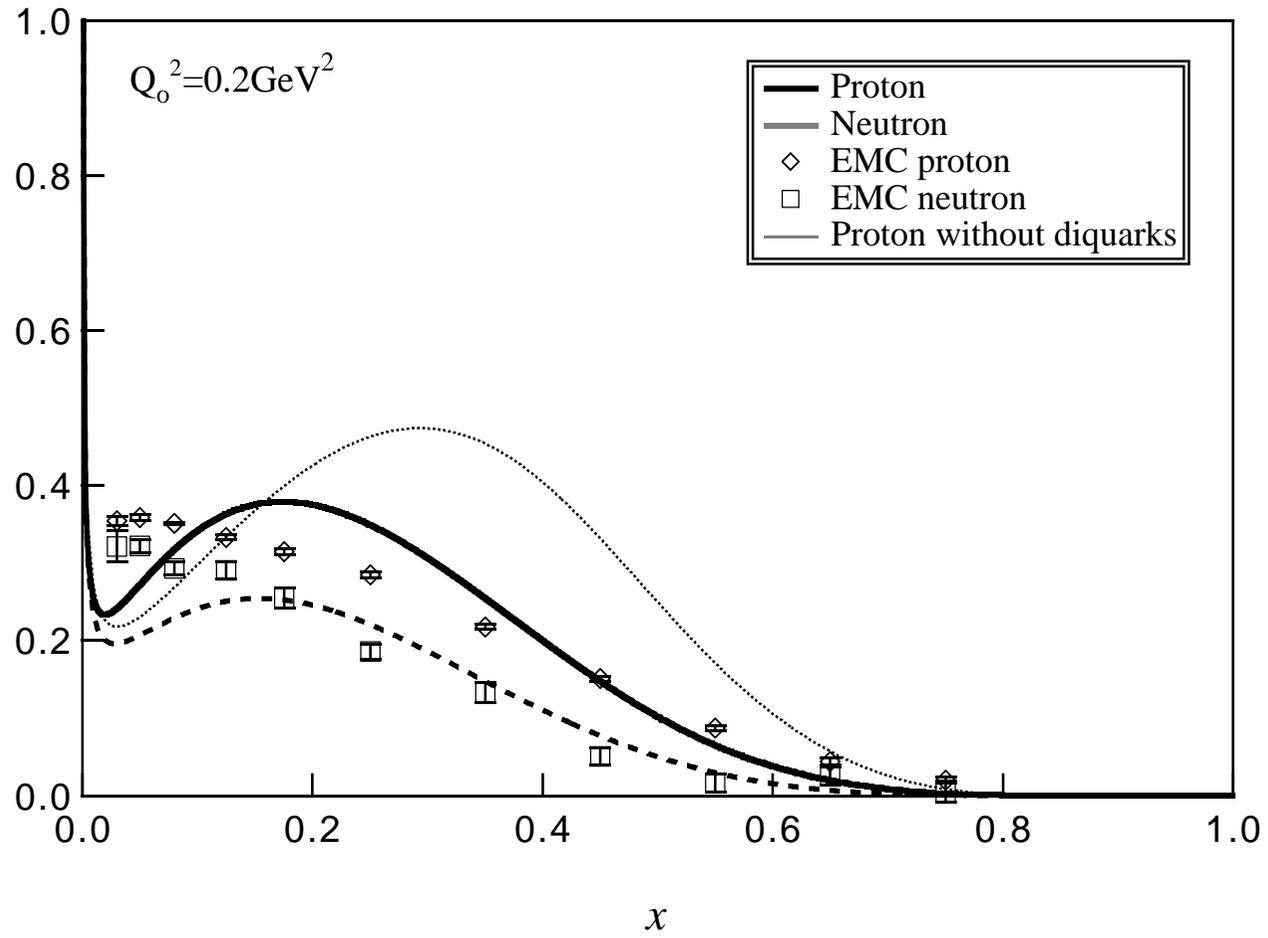
Fig. 3	: The proton and neutron structure functions at 
\(Q^{2}=15GeV^{2}\).  The solid curve represents  \(F_{2}^{p}(x)\), and 
the dashed curve  \(F_{2}^{n}(x)\).  The proton structure function 
calculated without diquark scattering processes is depicted by the 
dotted curve.  Here, we use the low energy model scale 
\(Q_{o}^{2}=0.2GeV^{2}\).  The experimental data are taken from the 
EMC experiment\cite{EMC}.  
\vglue 0.5cm

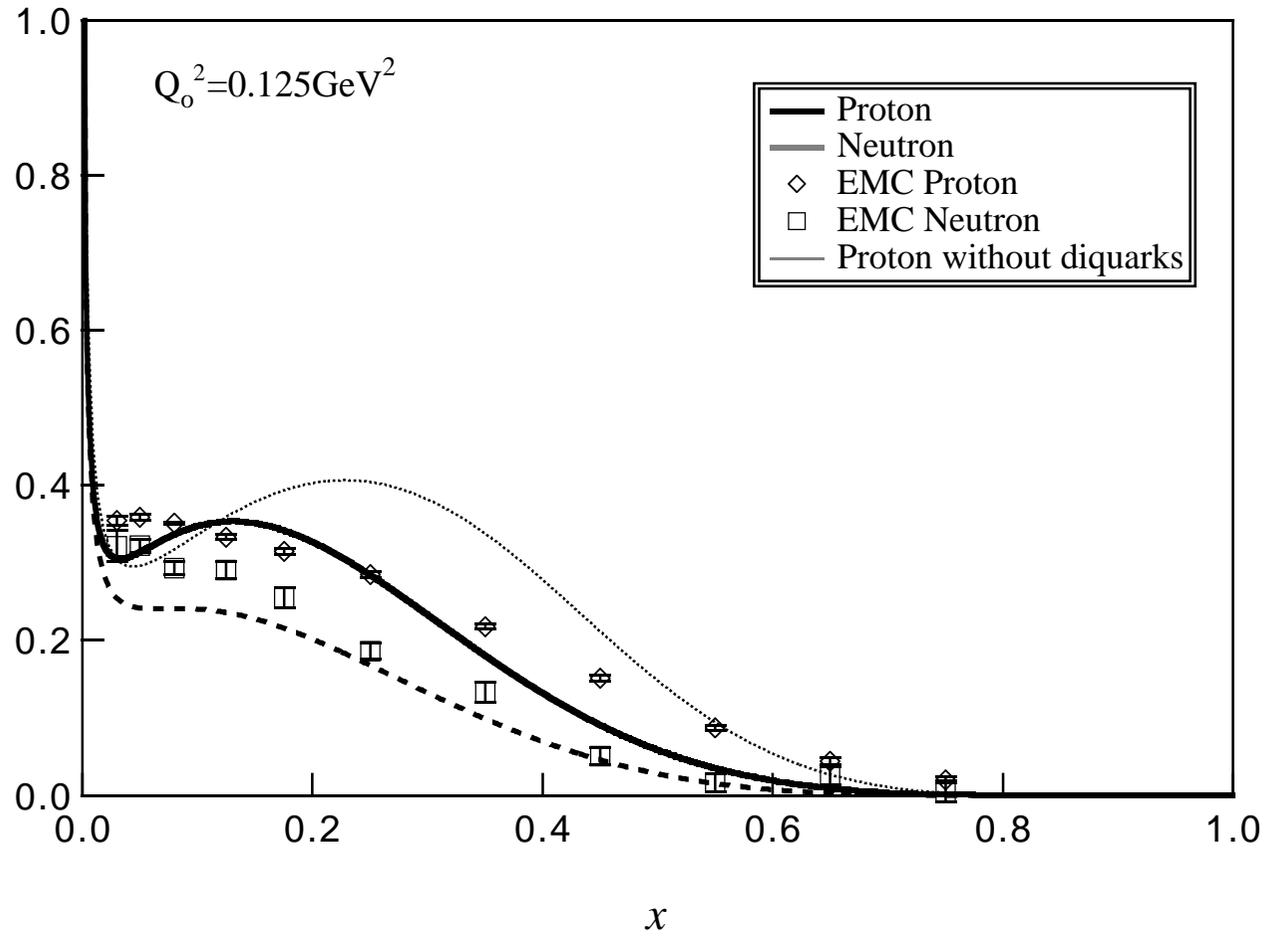
Fig. 4	: The proton and neutron structure functions at \(Q^2=15GeV^2\).  
The notations are the same as those of Fig. 3.  The low energy scale 
\(Q_{o}^{2}=0.125GeV^{2}\) is used.
\vglue 0.5cm

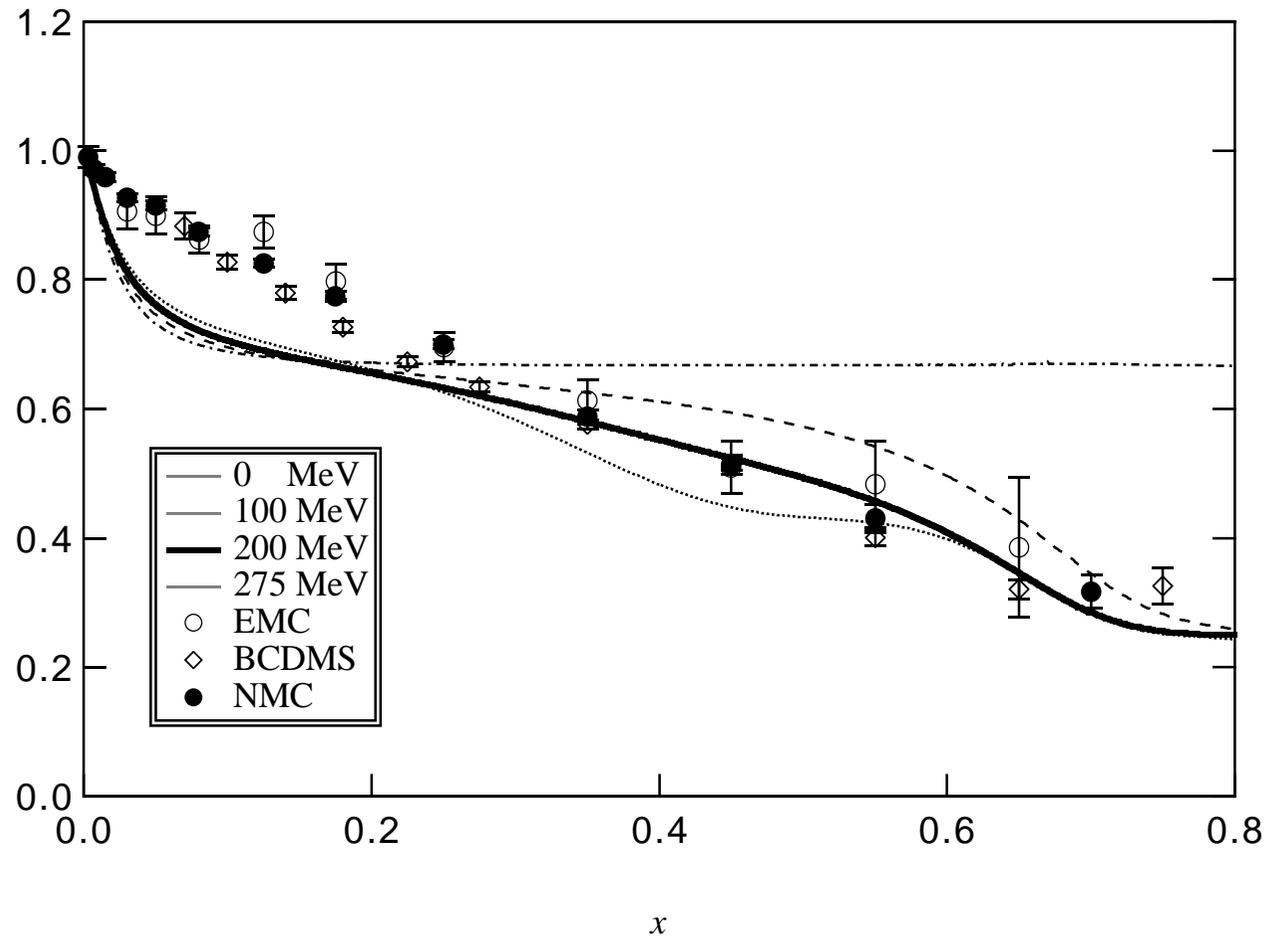
Fig. 5	: The ratio of the nucleon structure functions 
\(F_{2}^{n}/F_{2}^{p}\) at \(Q^{2}=15GeV^{2}\) with \(m_{A}-m_{S}=0, \, 
100, \, 200, \, 275MeV\) (\(Q_{o}^{2}=0.2GeV^{2}\)).  
The experimental data are taken from refs. [1-3].
\vglue 0.5cm

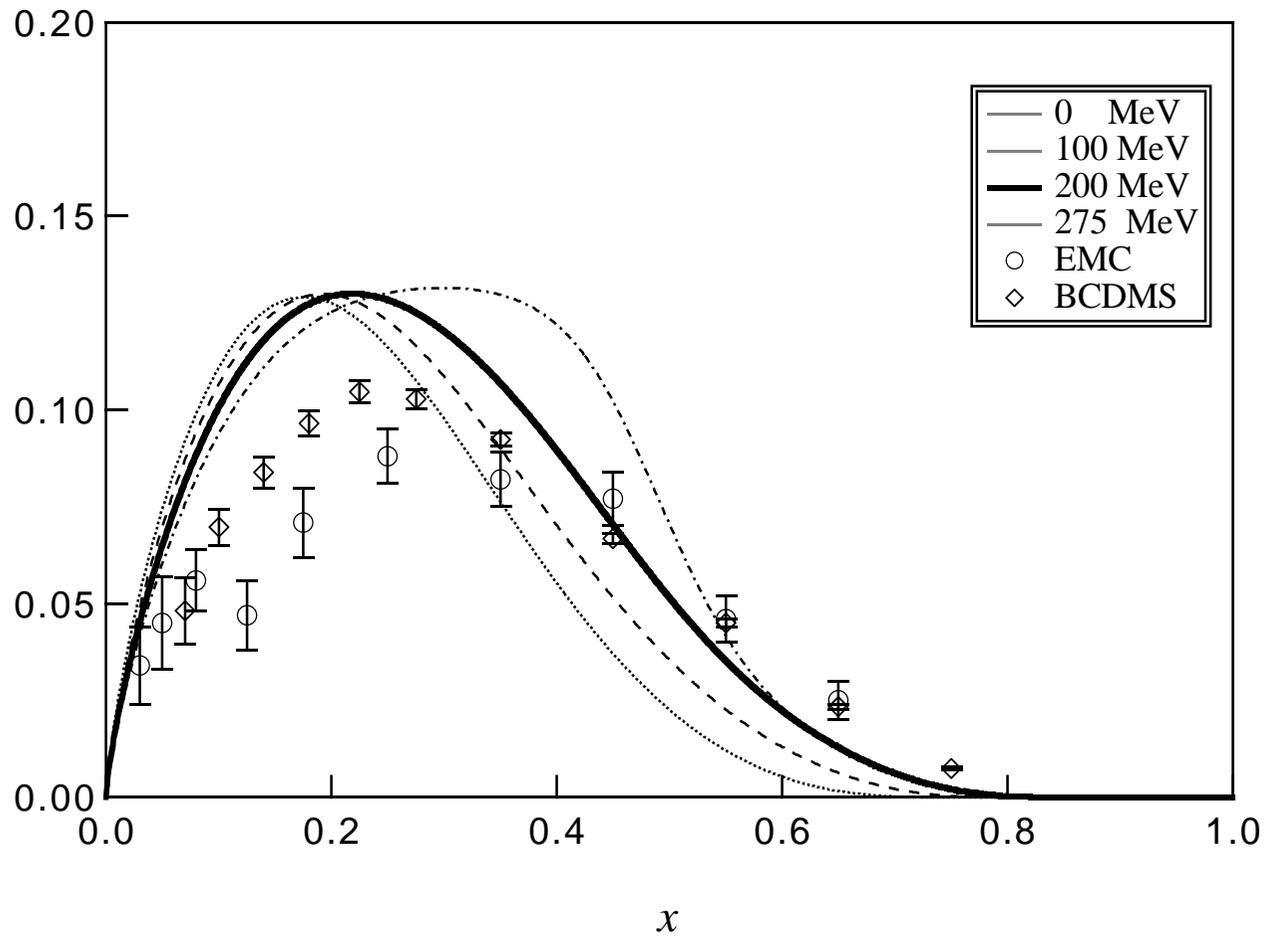
Fig. 6	: The difference of the nucleon structure functions 
at \(Q^{2}=15GeV^{2}\) with \(m_{A}-m_{S}=0, \, 100, \, 200, \, 
275MeV\) (\(Q_{o}^{2}=0.2GeV^{2}\)).  The experimental values with 
error bars are taken from the EMC and BCDMS experiments\cite{EMC,BCDMS}.
\end{document}